# Making AI Functional with Workarounds: An Insider's Account of Invisible Labour in Organisational Politics

**Research-In-Progress**

**Shang Chieh Lee**
School of Computer Science
University of Technology Sydney
Ultimo, New South Wales
Email: shangchieh.lee@uts.edu.au

**Bhuva Narayan**
School of Communication
University of Technology Sydney
Ultimo, New South Wales
Email: bhuva.narayan@uts.edu.au

**Simon Buckingham Shum**
Connected Intelligence Centre
University of Technology Sydney
Ultimo, New South Wales
Email: Simon.BuckinghamShum@uts.edu.au

**Stella Ng**
Management Discipline Group
University of Technology Sydney
Ultimo, New South Wales
Email: Stella.Ng@uts.edu.au

**A. Baki Kocaballi**
School of Computer Science
University of Technology Sydney
Ultimo, New South Wales
Email: baki.kocaballi@uts.edu.au

## Abstract

Research on the implementation of Generative Artificial Intelligence (GenAI) in higher education often focuses on strategic goals, overlooking the hidden, and often politically charged, labour required to make it functional. This paper provides an insider's account of the sociotechnical friction that arises when an institutional goal of empowering non-technical staff conflicts with the technical limitations of enterprise Large Language Models (LLMs). Through analytic autoethnography, this study examines a GenAI project pushed to an impasse, focusing on a workaround developed to navigate not only technical constraints but also the combined challenge of organisational territoriality and assertions of positional power. Drawing upon Alter's (2014) theory of workarounds, the analysis interprets "articulation work" as a form of "invisible labour." By engaging with the Information Systems (IS) domains of user innovation and technology-in-practice, this study argues that such user-driven workarounds should be understood not as deviations, but as integral acts of sociotechnical integration. This integration, however, highlights the paradoxes of modern GenAI where such workarounds for "unfinished" systems can simultaneously create unofficial "shadow" systems and obscure the crucial, yet invisible, sociotechnical labour involved. The findings suggest that the invisible labour required to integrate GenAI within complex organisational politics is an important, rather than peripheral, component of how it becomes functional in practice.

**Keywords** Autoethnography, Invisible Labour, Organisational Politics, Sociotechnical Systems, User Innovation, Workaround

# 1 Introduction

As Generative Artificial Intelligence (GenAI) is integrated into higher education, a gap can emerge between its technological potential and institutional realities. This study explores the challenges of making GenAI functional within existing organisational structures. It focuses on how users navigate the friction that arises between advanced systems, organisational constraints, and interpersonal dynamics.

## 1.1 Background and Problem Statement

University professional staff, positioned in the "third spaces" between academic and administrative domains (Veles et al., 2023, p. 147), play an important role in this process. Their experiences with GenAI can offer valuable insight into how systems are adapted in practice. These efforts can manifest as "workarounds" (Alter, 2014, pp. 1042-1045). Such adaptations can suggest that information systems are "unfinished" (Ciborra, 2002, p. 3), requiring significant "articulation work" to fit technologies into practice (Alter, 2014, pp. 1045, 1049, 1055; Gasser, 1986, p. 211; Strauss, 1985, pp. 8-9), a form of "invisible labour" that often extends beyond technical tasks to navigating organisational politics. Alter's (2014, pp. 1044, 1049) theory aligns with this view by framing user "adaptation" as a goal-driven process of overcoming system limitations, which, in the context of malleable GenAI, can be seen as a form of system completion. The flexibility of these systems suggests that user-driven innovation may be seen not as a deviation, but as a practice that fosters "worker-AI coexistence" by completing a system's design within a specific organisational context (Zirar et al., 2023, p. 7). This study's autoethnographic account of adapting malleable GenAI provides a detailed case relevant to IS research on technology-in-practice and user innovation, grounding these concepts in a real-world example of sociotechnical friction.

## 1.2 Research Gap and Approach

Analytic autoethnography provides the methodological approach to examine a workaround's development from an insider's perspective, as it captures the nuanced labour of navigating hierarchical power dynamics that other methods might not fully encompass. The approach combines personal narrative with workaround theory to analyse the interplay between user agency, system limitations, and the sociotechnical practices that make technology functional.

## 1.3 Research Focus and Questions

This inquiry is guided by the following research question: How might a user-driven workaround for a malleable GenAI system emerge and function amidst sociotechnical friction? To explore this, the study characterises the solution using an IS framework for workarounds before examining the resulting changes to work practices and the user-system relationship. It synthesises these findings to analyse the broader sociotechnical implications of integrating malleable GenAI into institutional workflows.

# 2 Theoretical Foundations

This study's theoretical foundation introduces the sociotechnical systems perspective, situates the research within IS literature on workarounds, and argues that malleable GenAI creates a new context for these adaptive practices.

## 2.1 A Sociotechnical Systems Perspective

A sociotechnical systems (STS) perspective provides the theoretical lens for this study, positing that work systems are composed of interdependent technical systems and the surrounding "social arrangements" (Ciborra, 2002, p. 3). Rather than viewing technology as a separate force, the STS lens emphasises that effective performance requires the "joint optimization" of both elements (Trist, 1981, p. 24). Within IS, this perspective evolved into structurationist views such as the "duality of technology," which frames technology and organisational practice in a recursive relationship of mutual constitution (Orlikowski, 1992, p. 406). Adopting an STS lens provides a framework for moving beyond a purely technical analysis to examine the workaround as an emergent practice arising from sociotechnical friction. This focus on the human and organisational dynamics, often overlooked in technology-centric narratives, makes the STS lens particularly useful for studying digital transformation.

## 2.2 Workarounds in Information Systems

Alter's (2014, pp. 1044, 1055-1058) process "theory of workarounds" provides the specific framework for this analysis. A workaround is a user-driven solution to perceived system limitations (Alter, 2014, pp. 1042-1045), often developed to overcome organisational "obstacles" (Alter, 2014, pp. 1048, 1050). IS research has sometimes framed these practices as "non-compliant user behaviors" (Azad & King, 2008, p. 265) or as a “costly alternative” to the intended design of a rigid system (Petrides et al., 2004, p. 100). This perspective often highlights the friction when standardised systems meet the “messy” realities of organisational life (Ciborra, 2002, p. 26). This study builds on this understanding by exploring how workarounds are shifting in the context of malleable AI, using the framework to interpret the autoethnographic account and characterise its sociotechnical implications.

### 2.3 Malleable GenAI and the Context for Workarounds

In contrast to workarounds for rigid systems, the context of malleable GenAI offers a different perspective. Foundation models are considered “unfinished” by design (Bommasani et al., 2022, pp. 3, 7, 9, 17), and this inherent malleability provides a basis for viewing user adaptation not as a deviation, but as a potentially valuable practice. This practice-based perspective aligns with the foundational concept of the “duality of technology” (Orlikowski, 1992, p. 406). Such practices can therefore be understood as a form of “technologies-in-practice” (Orlikowski, 2008, p. 262) that complete a system’s functionality, fostering a “symbiotic relationship” that enables “worker-AI coexistence” (Zirar et al., 2023, p. 7). Therefore, this study explores GenAI workarounds as a potential mechanism for sociotechnical integration, rather than as instances of user non-compliance. However, this same malleability can also create new arenas for sociotechnical friction, as stakeholders must then negotiate the technology's role and implementation within established workflows.

## 3 Research Approach and Method: Analytic Autoethnography

To ground IS insights in personal experience, I use analytic autoethnography. I selected this method because the phenomenon of interest, the nuanced and often “invisible labour” of navigating sociotechnical friction, is difficult to capture through external observation alone. The approach is characterised by the researcher being a “complete member researcher” in the research setting, visible in the text and committed to developing theoretical understandings of social phenomena (Anderson, 2006, p. 378). Unlike evocative autoethnography’s focus on storytelling, this analytic approach develops theoretical explanations of “broader social phenomena” (Anderson, 2006, p. 375). This approach aligns well with IS research traditions that position story and theory symbiotically, where story illustrates “personal nuances” and theory explains them to a wider audience (O’Raghallaigh et al., 2016, p. 2). The method can be useful for understanding the “world of the ‘user’,” especially with new technologies where usage patterns are not yet established (Hardwicke & Riemer, 2018, p. 2).

### 3.1 Key Features of Analytic Autoethnography

Applying Anderson's (2006, p. 378) “five key features of analytic autoethnography” involved being a “complete member researcher” with firsthand access to the setting, practicing “analytic reflexivity” through critical self-interrogation, attending to the “narrative visibility of the researcher’s self” in the text, engaging in a “dialogue with informants beyond the self” to refine insights, and maintaining a firm “commitment to theoretical analysis.” This self-interrogation is further conceptualised as “critically engaged care of the self” (Huber, 2022, p. 5). Accordingly, the study's analytic goal is “to use empirical data to gain insight into some broader set of social phenomena than those provided by the data themselves” (Anderson, 2006, p. 387).

### 3.2 Data Collection and Analysis

The data for this study primarily comprises my own “lived experiences,” recorded as “field notes” (Kaltenhauser et al., 2024, pp. 9, 12) that contain “embodied sensations, emotional reactions, and critical reflections” (Tarisayi, 2023, p. 57). I also used retrospective data from artifacts such as emails to revisit past experiences. This method allowed me to deploy external sources to “augment my personal recollection and give a depth and richness to the narrative” (Lee, 2019, p. 1). My analysis involved identifying recurring themes and “tacking back and forth between empirical materials and emergent theories” to construct patterns that helped to explain the workaround (Huber, 2022, p. 6).

### 3.3 Ethical Considerations and Researcher Vulnerability

In conducting this autoethnography, I navigated what Sparkes (2024, p. 107) calls an “ethically contested terrain,” a process guided by relational ethics that require “accountability and care” when representing others (Edwards, 2021, p. 2). A primary ethical challenge stems from the “personalised nature of autoethnography,” which “means that other parties can be implicated in the research” (O’Raghallaigh et al., 2016, p. 5), intertwining personal and professional relationships. This concern is heightened when using external data sources, as an autoethnography often has at its core “the behaviour of others and my relationships with them” (Lee, 2019, p. 7). A central task, then, involves navigating this ethical terrain carefully, guided by the principle that "it is important to consider seeking process consent and protecting the privacy and identities of participants" whenever other people are included in the narrative (Kaltenhauser et al., 2024, p. 15). One common response to this challenge is the anonymisation of names. This practice can itself be a source of anxiety for the researcher; one autoethnographer, for example, described feeling "vaguely inauthentic by using an alternate name" when presenting her work (Wall, 2008, p. 49). Despite this personal tension, anonymisation serves to "constitute" critical "knowledge" while protecting those involved (Huber, 2022, pp. 12-13). In the following autoethnographic account, all names of colleagues and specific role identifiers have been changed or removed. Furthermore, autoethnography entails an "ethic of the self," which acknowledges the researcher's vulnerability (Edwards, 2021, pp. 3-4). The decision to proceed, therefore, involves weighing the potential "benefits" of the research against these personal costs. This calculation represents a key part of the autoethnographic process itself (Kaltenhauser et al., 2024, p. 15), particularly when the research aims to give a "new voice" to previously unrepresented experiences (Lee, 2019, p. 8).

# 4 An Autoethnographic Account of the Workaround

The following autoethnographic account provides the empirical basis for this study, detailing the technical barrier, sociotechnical friction, and the adaptive solution created to make the system function.

## 4.1 The Catalyst: A Sociotechnical Friction Point

The project aimed to empower a non-technical team by building a conversational interface for Staff Development Fund (SDF) data, using Microsoft Copilot to query a large transaction log. However, testing revealed that the full transaction log, projected to exceed 35,000 rows annually for a single faculty, would surpass the GenAI's context window. This experience exemplifies a documented challenge in applying LLMs to tabular data, where expansive grids create token-inefficient inputs that exceed model limitations. This constraint is amplified when "noisy information becomes an issue in large tables for LMs" (Sui et al., 2023, as cited in Fang et al., 2024, p. 10) and requires careful management of dataset size; for example, one study "included datasets with... at most 30 columns to stay within T0's token limit" (Hegselmann et al., 2023, p. 4). Furthermore, even if a table fits the context window, a separate challenge arises. Research suggests that LLM performance is "highest when relevant information occurs at the beginning or end of the input context and significantly degrades when models must access relevant information in the middle of long contexts" (Liu et al., 2024, p. 157). Consequently, even advanced models with large context windows (Microsoft, 2025) would likely be insufficient to overcome these token limits and performance degradation. This technical barrier prevented the AI from processing the entire dataset, creating a "perceived need for a workaround" and rendering the initial plan unworkable (Alter, 2014, p. 1057). This limitation suggests a gap between the foundation model's general-purpose nature and the enterprise's large-scale data needs, which highlights the system's "unfinished" quality.

## 4.2 A Human-in-the-Loop Solution

The project stalled because the conversational AI, though suitable for users, could not process the large dataset. To resolve this, I developed a two-part workaround to automate data processing, using a Python script to join and summarise the raw data into digestible report files for the custom Copilot. This process transformed an unsuitable design into a functional sociotechnical system. Reshaping the data to fit the AI's constraints constituted "articulation work." Strauss (1985, p. 8) described this concept as a "supra-type of work" involving "the meshing of the often numerous tasks, clusters of tasks, and segments of the total arc." Gasser (1986, p. 211) later specified this for computing environments as the work that "serves to establish, maintain, or break the coordinated intersection of task chains". This technical adaptation, however, was only one part of the effort.

## 4.3 The Labour of Friction: Navigating Organisational Politics

While the technical work was largely unseen, the more significant "invisible labour" involved navigating the project's human and political friction. I managed this social friction by addressing unspoken concerns about job redundancy, focusing on transforming processes to augment, rather than replace, the core responsibilities of others. A more direct challenge involved navigating what I perceived as a pattern of organisational politics. This began with a manager, an informal expert on IT tools, who made a suggestion on a prior project that I concluded was an ostensibly helpful attempt to integrate the solution with a centrally-managed system and mire the project in bureaucracy. My decision not to adopt this was later met with what I perceived as an escalation to assertions of positional power, with the manager publicly questioning my formal job title in what I interpreted as a tactic to diminish my professional standing. This was compounded by a senior leader imposing prohibitive, zero-cost conditions and mandating a centralised approval process to control the staff development initiative. This experience highlighted for me the hidden emotional labour required to navigate the political and bureaucratic landscape and ultimately make the workaround functional. This political impasse prompted my strategic shift to pursue the project through a more collaborative pathway with academics from other faculties and a pan-university research centre.

# 5 Discussion: A Theoretical Analysis of the Workaround

To characterise the workaround as a "goal driven adaptation" (Alter, 2014, p. 1044), I analyse my autoethnographic account through Alter's "process theory" (p. 1055). This theory offers a "structural foundation" for tracing a workaround's development, supported by subsequent IS research (Wibisono et al., 2019, p. 188).

## 5.1 Characterising the Workaround with Process Theory

Applying Alter's (2014) process theory, I use the operational model from Wibisono et al. (2019) to structure my characterisation through its distinct developmental phases, terms, and abbreviations. The process began with conflicting Intentions, Goals, and Interests (IGI), as the goal of an inclusive, AI-powered query system clashed with the technical limitations of the AI model, specifically its context window. This conflict created a Perceived Need for a Workaround (PNW). In the subsequent cognitive phases of Identification of Possible Workarounds (IPW) and Selection of Workaround to Pursue (SWP), autoethnography captures the internal monologue moving from "The AI cannot do this" to "What if I pre-process the data for the AI?". This cognitive leap can be interpreted as a human-in-the-loop decomposition of the task. The final phase, Development and Execution of Workaround (DEW), involved "bricolage", the process of creating something from available resources (Lévi-Strauss, 1967, as cited in Alter, 2014, p. 1049), which manifested as writing the Python script. To ground this case, the workflow aligns well with a generalised model from Fang et al. (2024, p. 8) for leveraging LLMs with tabular data (see Appendix 1). The Python script performs the "Serialization" and "Table Manipulations" to create a summarised input for the "LLM Agent". This comparison leads me to suggest that the adaptive actions in this case, rather than being idiosyncratic, may align with recognised practices for handling tabular data with LLMs.

## 5.2 The Paradoxes of the Workaround: Innovation and Shadow IT

This solution, however, can be interpreted as introducing a set of interconnected paradoxes. While the workaround solved a technical problem, its necessary development in private created shadow IT and obscured the sociotechnical labour required for its creation. The automated process itself can be seen as a "goal-driven adaptation" (Alter, 2014, p. 1044) that suggests user agency, and in the context of malleable GenAI, such adaptations may be viewed not just as solutions, but as necessary practices that foster "worker-AI coexistence" (Zirar et al., 2023, p. 7) and function as a form of "technologies-in-practice" (Orlikowski, 2008, p. 262). The workaround resulted in a system that can "replicate in full or in part data and/or functionality of the legitimate systems of the organization" but operates outside official governance (Behrens & Sedera, 2004, p. 1713). This unofficial "shadow system" (Alter, 2014, p. 1046) highlights the dual nature of workarounds as "both inventive solutions to pressing organizational needs and over time, and costly alternative to a robust and flexible information system" (Petrides et al., 2004, p. 100). This tension seems pronounced with malleable GenAI, where user-driven solutions address immediate problems while creating shadow systems that obscure underlying limitations. Unlike traditional shadow IT that often supplants official systems, this workaround completes a system that is unfinished by design. This compounds these paradoxes, as the workaround in turn reinforces the technical dependency it aimed to eliminate while its "shadow" nature obscures the required sociotechnical labour.

## 5.3 Surfacing Invisible Labour as Articulation Work

These paradoxes of the workaround extend beyond the creation of shadow IT to encompass the hidden sociotechnical labour of its integration. This "invisible labour" can be understood as a form of "articulation work" (Strauss, 1985; Gasser, 1986). An important part of this work involved countering what I perceived as a coordinated pattern of organisational politics, which manifested as both organisational territoriality and assertions of positional power. This labour also extended to mitigating social risks among colleagues and overcoming external bureaucratic inertia, which in one instance blocked progress for months until a senior academic intervened. The "unfinished" nature of malleable GenAI (Bommasani et al., 2022, pp. 3, 7, 9, 17) may amplify this form of labour by expanding the scope for negotiation over the technology's use and purpose. Managing the vertical power structure was particularly demanding. The interactions escalated from what I perceived as ostensibly helpful suggestions designed to create bureaucratic traps to direct assertions of hierarchical status. One manager engaged in professional gatekeeping, a common expression of organisational territoriality, by making public challenges to my professional standing. Based on a prior project, the manager also attempted to steer our work into bureaucratic traps, seemingly to protect his established domain of technical expertise against a more advanced approach. A senior leader, in turn, exercised institutional authority by imposing restrictive conditions and centralising control over the initiative, reinforcing the existing hierarchy. This perspective views the workaround not as an isolated technical fix, but as a complex sociotechnical practice that demanded the emotional and political labour needed to maintain the "coordinated intersection of task chains" (Gasser, 1986, p. 211).

### 5.4 A Sociotechnical Integration Perspective on Workarounds

Building on this concept of articulation work, I propose viewing the workaround not as a deviation, but as an act of sociotechnical integration. The "unfinished" nature of AI foundation models suggests that user adaptation is a core feature of implementation. From this perspective, AI adoption may be viewed as a continuous process rather than a discrete technical event. This perspective aligns with a practice-based view of technology, where enacting "technologies-in-practice" (Orlikowski, 2008, p. 262) is an expression of the recursive relationship described in the "duality of technology" (Orlikowski, 1992, p. 406). The workaround can therefore be seen not as a deviation, but as a user-driven "technology-in-practice" that completes the system's functionality. With malleable technologies like LLMs, the line between use and design blurs, and practices such as the pre-processing and serialisation of tabular data can become central to the implementation process, rather than peripheral fixes (Fang et al., 2024, p. 8).

### 5.5 Implications for Theory and Practice

Organisations, for instance, may benefit from viewing workarounds as insights into sociotechnical friction, rather than simple user deviations. This perspective suggests a potential extension of workaround theory that frames the "problem" not as user non-compliance, but as a system's inability to accommodate work needs. This extension would also recognise that organisational politics can shape the form a workaround takes, influencing whether it becomes a piece of collaborative innovation or, as in this case, a form of shadow IT. This view suggests designing AI tools for malleability, establishing channels for user innovation that can make these adaptations visible and legitimate, mitigating the political risks that push such work into the shadows, and recognising "articulation work" (Gasser, 1986, p. 211) and the "supra-type of work" (Strauss, 1985, p. 8). This analysis suggests that in the context of malleable AI, user-driven workarounds might be viewed less as deviations and more as "technologies-in-practice" that help complete the sociotechnical system. This perspective aligns with viewing user activity not as non-compliance, but as a central component of "worker-AI coexistence" (Zirar et al., 2023, p. 7), especially when official tools lack needed flexibility. Understanding this shift may inform future GenAI system design and institutional governance that learns from these adaptations.

## 6 Conclusion

This study's analysis suggests user-driven workarounds emerge from the often "invisible labour" of navigating sociotechnical friction. The autoethnographic account indicates these adaptive practices function as an important form of user innovation, necessary to integrate malleable and inherently "unfinished" GenAI systems into practice. The insider perspective was particularly valuable for grounding this analysis in the real-world organisational politics of making GenAI functional.

### 6.1 Limitations and Directions for Future Research

While the findings of this single-case autoethnography are context-specific and not statistically generalisable, they suggest several avenues for future IS inquiry. Future research could explore the transferability of these findings, perhaps by utilising collaborative autoethnographies to "critically juxtapose their different life experiences" (Kaltenhauser et al., 2024, p. 2). Further inquiry could investigate the specific organisational and political conditions that influence whether user adaptations become visible, collaborative innovations or are pushed into the shadows as unsanctioned IT. Further avenues include exploring the long-term ethical impacts, such as the relational "ethic of the self" (Edwards, 2021, pp. 3-4) and the ethical "afterlife" of the publication (Sparkes, 2024, p. 128). Another direction involves leveraging autoethnography to examine emergent human-AI task allocations, as the user-led pre-processing in this case highlights the critical uncertainty where "we are still unsure what happens to worker-AI task allocations" (Zirar et al., 2023, p. 11). Studying these real-world adaptations can inform the design of more inclusive sociotechnical systems that value users' adaptive practices.

# Appendix 1

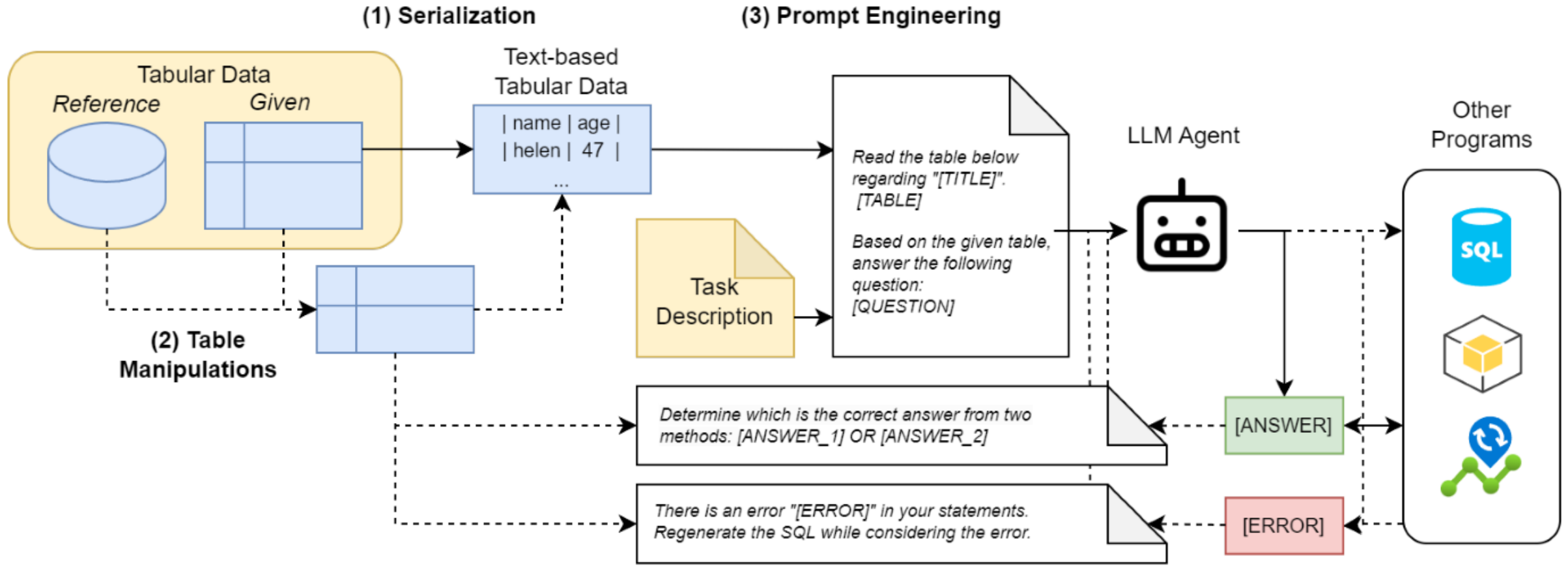

*Figure 1: Key techniques in using LLMs for tabular data. Source: Fang et al. (2024)*

# Acknowledgements

The work of the first author was supported by the Australian Government Research Training Program (RTP) Scholarship. The first author wishes to express sincere gratitude to the supervisors, Bhuva Narayan, Simon Buckingham Shum, Stella Ng, A. Baki Kocaballi, for their invaluable guidance and support. The authors also thank their academic and professional colleagues for their constructive feedback and assistance throughout this research. Microsoft Copilot was utilised to improve the language and readability; the authors reviewed and edited all suggestions and retain full responsibility for the final content.

# Copyright